\documentstyle[nanosymp,epsf]{article}
\begin{document}

\title{\centering{Effect of Screening on Spin Polarization in a Two-Dimensional Electron Gas}}
\author{\centering{V.~V'yurkov$\dag$ and A.~Vetrov$\ddag$}}
{$\dag$Computer Solid State Physics Laboratory, University of Aizu, \\Aizu-Wakamatsu 965-8580, Japan\\
$\ddag$Institute of Physics and Technology, Russian Academy of Sciences, \\Moscow 117218, Russia}
\footnote{$\dag$ 
Corresponding author. 
E-mail: vyurkov@u-aizu.ac.jp\\
On leave from Institute of Physics and Technology, Russian Academy of Sciences,
Moscow 117218, Russia
}

\beginabstract

The experimentally observed amazing dependence of a critical magnetic field $B_{c}$ of a full 
field-induced  
spin polarization and a spin susceptibility $\chi^{*}$ of a two-dimensional electron gas 
on the electron density
can be explained by screening of the Coulomb potential. 
The possibility of spontaneous full spin polarization expected at lower electron densities 
also crucially depends on screening.

\endabstract

\section{Introduction}

Among spin related phenomena one of the most intriguing is a spontaneous spin polarization (magnetization) 
of a two-dimensional electron gas (2DEG). 

In absence of direct methods to measure a 2DEG magnetization 
only indirect evidence is so far available. 
The experiments of Shashkin et al. [1] revealed 
almost a linear dependence of a critical magnetic field $B_c$ 
of full spin polarization versus 2DEG density $n$ in Si inversion layers. 
The full polarization state was associated 
with the onset of 
a saturation of a magnetoresistance in a parallel magnetic field. 
The extrapolation of the experimental curve to $B_c$=0 gave a non-zero value of $n$.
It looked like a manifestation of a possibility of spontaneous spin polarization at lower electron density. 
However, the recent breakthrough to a more dilute 2DEG allowed 
to Pudalov et al. [2] to discover 
a substantial declination of $B_c(n)$ curve from the conjecture of Ref.~[1]. A possibility 
of spontaneous spin polarization remains thus still unclear [3,4].   

The main goal of present communication is to elucidate the impact 
of screening on a 2DEG spin polarization. 
Just screening results in the behavior of $B_c(n)$ and $\chi^{*}(n)$ 
similar to that in the experiment.  
\section{Model of induced and spontaneous spin polarization}
We adopt a simplified model for the two-dimensional Fourier transform 
of the screened Coulomb potential  
\begin{equation}\label{eq1}
V(q) = \frac{2 \pi e}{\kappa} \frac{1}{q+ \lambda^{-1} },
\end{equation}
where $\kappa$ is the permittivity, $e$ is the elementary charge, and 
$\lambda$ is the model screening length. The latter could be defined 
by the 2DEG itself or by adjacent electrodes.

The polarization degree $\eta = |n_{+} -n_{-}|/n$ where $n_{+}$ and $n_{-}$ are
spin-up and spin-down electron densities, respectively, 
while $n=(n_{+} +n_{-} )$ being the total electron concentration 
is derived 
via minimization of a total energy $E_{tot}$ including kinetic, 
interaction (exchange, correlation), and Zeeman contributions. 
Here the direct Coulomb interaction is omitted as it does not 
depend on a spin polarization.
We employ a quite common model   
\begin{equation}\label{eq2}
E_{tot}= \frac{\pi \hbar^2}{g_{\nu}m_{b}} (n_{+}^{2} + n_{-}^{2}) + 
I -
\frac{1}{2} g_0 \mu _B B (n_{+} - n_{-}),
\end{equation}
where $I$ is an interaction term, 
$m_{b}$ is the mean electron band mass, 
$g_{\nu}$ is the valley degeneracy, 
$g_{0}$ is the free electron g-factor, 
$\mu_{B}$ is the Bohr magneton, 
and $B$ is a magnetic field. 
The valley degeneracy $g_{\nu}$ equals 4 for the 2DEG in a silicon structur. 

For the non-interacting electron gas ($I$=0) the minimization 
of $E_{tot}$ with respect to a polarization degree $\eta$ gives rise 
to a linear dependence of $B_{c}$ on $n$. Obviously, $B_{c}$=0 when 
$n$=0. As for the spin susceptibility $\chi^{*} \propto \eta/B$, 
it is constant and proportional to the product $m_{b}g_{0}$=$2m_{b}$. 
For interacting electron gas it is convenient to express 
the spin susceptibility as $\chi^{*} \propto m^{*}g^{*}$ 
where $m^{*}g^{*}$ is a virtual product of electron mass and g-factor.  
 
For fairly small electron densities the interaction term in Eq.~(2) 
becomes essential. 

Two plausible models of interaction could be used. 
The first one is that of Kohn-Sham (KS), widely employed 
for the interacting 2DEG description. 
The second one invokes the Ising (I) model
widely exploited for spin lattice description. 
According for the KS model the interaction is
\begin{equation}\label{eq3}
I_{KS}= I(n_{+}) + I(n_{-}).  
\end{equation}
Here the exchange and correlation belong to the electrons with parallel 
spins.
If the interaction is associated with exchange, 
the values of $I(n_{\pm})$ are supplied by integrals
\begin{equation}\label{eq4}
I(n_{\pm}) = - \frac{g_{\nu}}{2} \int_{\Omega_{\pm}}\frac{d^{2}k_{1}}{(2\pi)^{2}}
\int_{\Omega_{\pm}}\frac{d^{2}k_{2}}{(2\pi)^{2}}
eV(k_{1}-k_{2}),   
\end{equation}
where the integrations are performed over the circle interior $\Omega_{\pm}$
corresponding to the Fermi wave vector $k_{F \pm}= \sqrt{\pi n_{\pm}}$ (see Fig.1).
In the limit $k_{F\pm} >> 1/ \lambda$ 
\begin{equation}\label{eq5}
I(n_{\pm}) \sim -(e^2/\kappa k_{F\pm}) n_{\pm}^2 .  
\end{equation}
i.e., as that for unscreened potential. 
In the opposite limit when $k_{F\pm} >> 1/\lambda$ 
\begin{equation}\label{eq6}
I(n_{\pm}) \sim - (e^2 \lambda/ \kappa) n_{\pm}^{2}.  
\end{equation}

\begin{figure}[h]
\leavevmode
\centering{\epsfbox{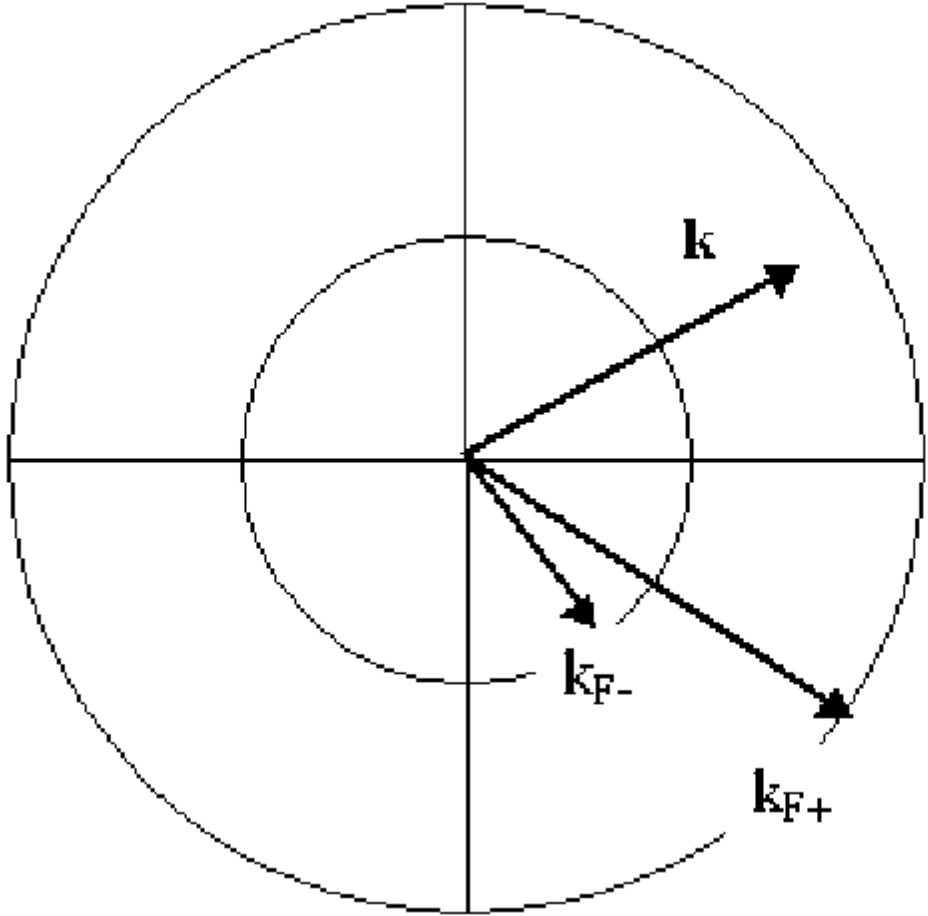}}
\caption 
{ 
Two-dimensional momentum k-space of electrons in 2DEG; 
$k_{F+}$ and $k_{F-}$ denote the corresponding Fermi 
wave vectors for spin-up and spin-down polarized electrons, respectively.  
}
\label{fig1}

\end{figure}

In the intermediate range of $k_{F\pm}$ 
the interaction terms $I(n_{\pm})$ may be numerically calculated via Eq.~(4)
for the screened Coulomb potential of Eq.~(1). 
Evidently, the critical field $B=B_c$ should be associated with $\eta=1$. 
According to the Ising model the interaction term looks like
\begin{equation}\label{eq4}
I_{I} = I_{KS} +
g_{\nu} \int_{\Omega_{+}}\frac{d^{2}k_{1}}{(2\pi)^{2}}
\int_{\Omega_{-}}\frac{d^{2}k_{2}}{(2\pi)^{2}}
eV(k_{1}-k_{2}).   
\end{equation}
Here the interaction among electrons with opposite spin orientation is added. 

\section{Spin polarization and conductivity}
The interplay 
of spin polarization and conductivity is worthy of being touched 
here as the experiments mentioned dealt with measurements of conductivity 
to detect the spin polarization. 

Actually, a probable reason of the resistance 
growth and saturation observed in the experiments might be 
a lowered density of states (DOS) at the Fermi level caused by spin polarization. 
It could be also a likely cause of the metal-insulator transition if 
a spontaneous spin polarization (full or partial) may occur at zeroth magnetic field.
Theoretically the lowered DOS at the Fermi level in the spin polarized state 
was derived for 1DEG [5] to explain the observable declination 
of a quantum wire conductance from the conductance quantum, in particular, 
the so called "0.7 step". 
The existance of the lowered DOS ("pseudo-gap") at the Fermi level
caused by spin polarization was also confirmed for 2DEG 
by calculations in [6]. 

Worth noting for the dependence of DOS deepening
on the spin polarization degree to be substantial
the Ising-like model for the exchange interaction 
should be employed instead of the conventional Kohn-Sham model.
In detail this problem is discussed in [7].

\section{Results and discussion}

The calculated dependences of the critical magnetic field $B_c$ on 
the electron sheet density $n$ 
for the silicon structure are presented in Fig.2. We used the Ising model 
for the exchange interaction given by Eq. (7). 
The parameters of the structure are as follows. 
The electron mass equals $m$=0.36$m_{0}$ 
where $m_{0}$ is the free electron mass,
$\kappa= (\kappa_{Si}+\kappa_{SiO_{2}})/2$= 7.8 
is the mean permittivity at the Si/SiO$_{2}$ interface, 
and the valley degeneracy is $g_{\nu}$=4.
The curve (1) corresponds to the model screening length $\lambda$=13nm. 
The filled dots are taken from the experiment of Ref.~[5]. 
The dashed line reproduces the  conjecture of Ref.~[1]. 
\begin{figure}[b]
\leavevmode
\centering{\epsfbox{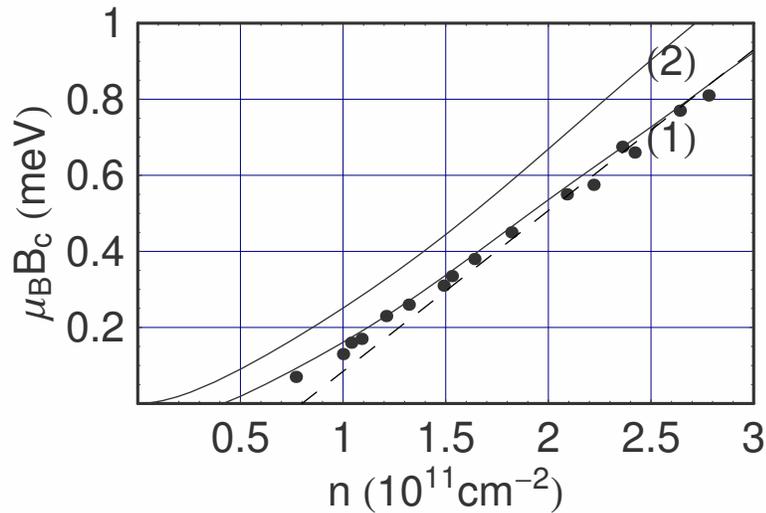}}
\caption 
{ 
The calculated dependences $\mu_B B_c$ vs electron sheet density $n$ (solid lines)
for screening length $\lambda=13nm$ (1) and $\lambda=5nm$ (2). The dots correspond 
to the experiment of Ref.~[5]. The dashed line reproduces the  conjecture of Ref.~[1].
}
\label{fig2}
\end{figure} 

The calculated curve (1) in Fig.2 to the best advantage coincides with the
experimental points. 
It predicts a spontaneous spin polarization for $n$=0.4 ~10$^{11}$~cm$^{-2}$ 
if the model screening length $\lambda$ fitted to experimental curve in Ref.~[4]
is actually as large as 13nm.
However, the achieved electron densities are above 0.7 ~10$^{11}$ cm$^{-2}$ yet. 
For fairly small screening length $\lambda$ (for instance, if it approaches the Bohr radius)
the screening results in both $B_c$ and $n$ simultaneously come to zero and 
spontaneous spin polarization does not exist.
This is illustrated by the curve (2) calculated for ($\lambda$=5~nm).

The calculated spin susceptibility $\chi^{*} \propto g^{*}m^{*}/2m_{b}$ against 
the parameter $r_{s}$ is plotted in Fig.~3. 
The ratio $r_{s} \sim (n)^{-1/2}/a_{B}$ characterizes the strength of 
Coulomb interaction with respect to the kinetic energy.
The calculated curve in Fig.~3 also agrees well with experimental 
points (filled circles) because it is merely another presentation of results in Fig.~2.

It should be emphasized that for unscreened Coulomb potential
both models for the exchange interaction (KS and I) presented here 
inevitably lead to the existence of spontaneous spin polarization at sufficiently low 
electron density, that is, at high value of the parameter $r_{s}$.
\begin{figure}[t]
\leavevmode
\centering{\epsfbox{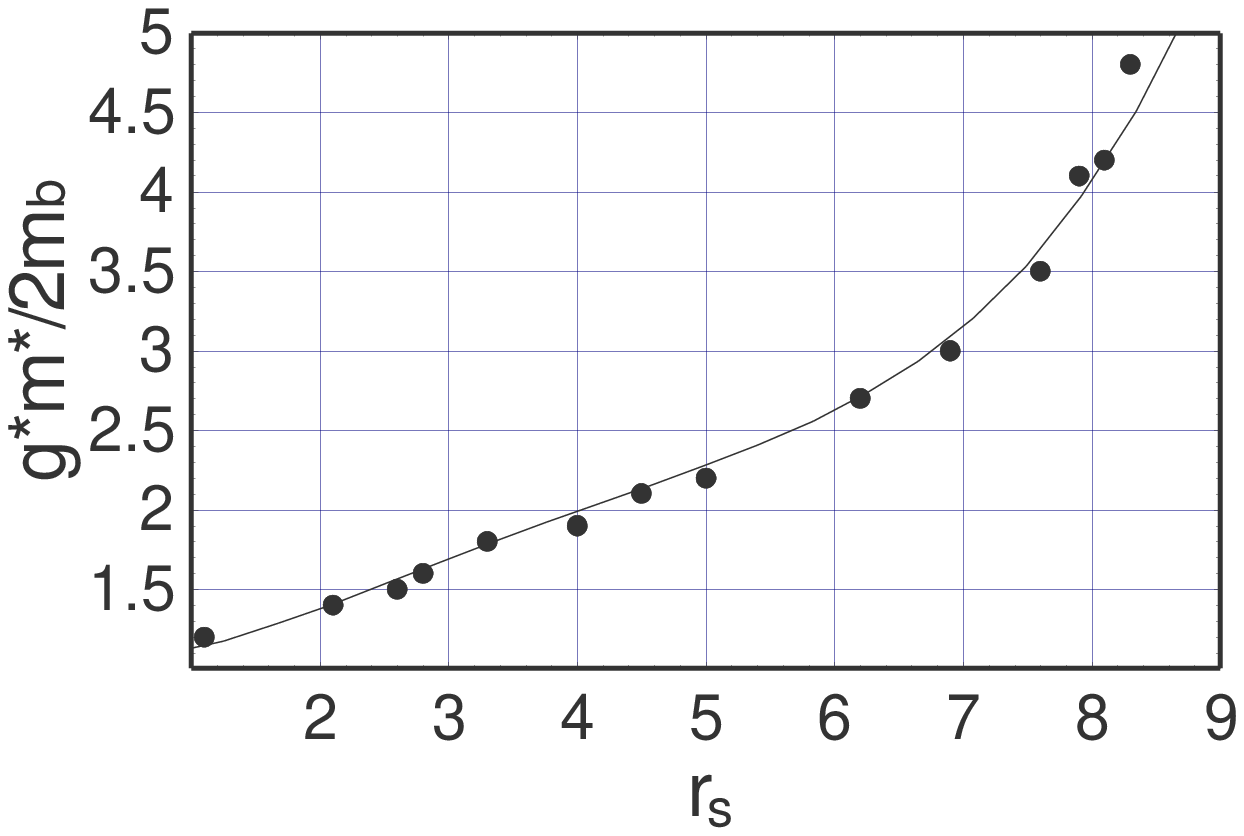}}
\caption 
{ 
The calculated dependence of the parameter $g^{*}m^{*}/2m_{b}$ 
proportional to the spin susceptibility $\chi^{*}$ against the parameter $r_{s}$
for screening length $\lambda$=13~nm. The dots correspond to the 
experimental data from Ref.~[2] 
}
\label{fig3}
\end{figure}  

The KS model was also used to fit to the experimental curve $B_{c}(n)$. 
The model screening length $\lambda$ turned out to be very large (about 100nm) 
compared with that obtained for the Ising model. It could be said that 
the Ising model is more "resistive" against screening.

As for the partial spontaneous spin polarization, i.e., ~$\eta < $  1, $B$=0, 
in the frame of the present models it could be hardly possible. 
Unlike to the induced spin polarization the spontaneous one has only 
an abrupt transition to a fully polarized state at sufficiently low electron density.

In the experiments [1-4] the critical field $B_{c}$, when 
the full spin polarization arose, was indirectly detected  
by the onset of 
a saturation of a magnetoresistance.
Therefore, it could be a discrepancy 
between the measured $B_{c}$ and the actual one.  
In our opinion, the critical field $B_{c}$ was underestimated in the experiments.
  
Hopefully, the methods of a direct measurement of 2DEG magnetization 
will be involved in the investigation. For example, 
in the recent experiments of Bagraev et al. [8] 
a nucleus magnetic resonance (NMR) of $^{29}$Si atoms 
in the vicinity of a quantum wire was employed to detect its spin polarization. 
It was the first direct confirmation of the assumption
that the so called "0.7 step" in a quantum wire conductance
originates just in a spontaneous spin polarization.  

\section{Conclusions}
The experimentally observed dependences of a critical magnetic field $B_{c}$ of a full 
field-induced  
spin polarization and a spin susceptibility $\chi^{*}$ of a two-dimensional electron gas 
on the electron density
are explained by screening of the Coulomb potential. 
The feasibility of spontaneous full spin polarization expected at lower electron densities 
also crucially depends on this screening.


\begin{thebibliography}{9}

 
%
\bibitem{1} 
A.~A.~Shashkin, S.~V.~Kravchenko, V.~T.~Dolgopolov, 
and T.~M.~Krapwijk, Phys. Rev. Lett. ~87  (2001) 086801.
%
\bibitem{2} 
V.~M.~Pudalov, M.~E.~Gershenson, H.~Kojima et al.,
Phys. Rev. Lett. ~88 (2002) 196404.
%
\bibitem{3} 
S.~V.~Kravchenko, A.~A.~Shashkin and V.~T.~Dolgopolov, 
Phys. Rev. Lett. ~89 (2002) 219701.
%
\bibitem{4} 
V.~M.~Pudalov, M.~E.~Gershenson, H.~Kojima et al., 
Phys. Rev. Lett. ~89 (2002) 219702.
%
\bibitem{5} 
V.~V'yurkov and A.~Vetrov, Nanotechnology 11 (2000) 336.
%
\bibitem{6} 
V.~V'yurkov and A.~Vetrov, cond-mat/0105361 (2001). 
%
\bibitem{7}
V.~V'yurkov, A.~Vetrov, and V.~Ryzhii, Int. Symp. Nanostructures: Physics and Technology,
St.Petersburg, June 23-28, 2003, Book of Extended Abstracts, to be published. 
\bibitem{8}
N.~T.~Bagraev, A.~D.~Buravleuv, W.~Gehlhoff et al., Physica E, to be published.
%
\newpage
\end{thebibliography}
\end{document}